\documentclass[10pt]{article}

\begin{document}

\title{Brane-Universes with Variable $G$ and $\Lambda_{(4)}$}
\author{J. Ponce de Leon\thanks{E-mail: jponce@upracd.upr.clu.edu}\\ Laboratory of Theoretical Physics, Department of Physics\\ 
University of Puerto Rico, P.O. Box 23343, San Juan, \\ PR 00931, USA} 
\date{Revised Version, August 2003}

\maketitle
\begin{abstract}
We investigate the cosmological consequences of a brane-world theory which 
incorporates time variations in the gravitational coupling  $G$  and the cosmological 
term $\Lambda_{(4)}$. 
We analyze  in detail the  model where $\dot{G}/G \sim H$ and
 $\Lambda_{(4)} \sim H^2$, which seems to be favored by observations. 
We show that these conditions single out models with flat space sections. 
We determine the behavior of the  expansion scale factor, as well as, the 
variation of $G$, $\Lambda_{(4)}$ and $H$ for different possible scenarios where 
the bulk cosmological constant, $\Lambda_{(5)}$,  can be zero, positive or negative. 
We demonstrate that the universe must recollapse, if it is  embedded in an Anti-de Sitter 
five-dimensional bulk, which is the usual case in brane models. 
We evaluate the cosmological parameters,  using some observational data,  
and show that we are nowhere near the time of recollapse. 
We conclude that the models with zero and negative bulk cosmological 
constant agree with the observed accelerating universe, while fitting simultaneously 
the observational data for the density and deceleration parameters. 
The age of the universe, even in the recollapsing case, 
is much larger than in the FRW universe. 

 \end{abstract}


\newpage
\section{Introduction}

In Einstein's General Relativity (GR) there are two fundamental physical parameters, 
the  gravitational coupling $G$ and the cosmological term $\Lambda_{(4)}$, which are 
typically assumed to be constants. However, the data emerging from the experimental 
and/or observational   verification of this assumption are not conclusive.  As a 
matter of fact,  recent measurements point toward the possibility that  small 
variations of these parameters cannot be excluded {\em a priori} \cite{Melnikov1}. 

The accuracy in the determination of $G$ is rather poor. Indeed, in the best case 
it does not exceed $10^{-4}$. Besides,  some experiments give inconsistent values 
for $G$, which do not overlap within their range of accuracy. Taking into consideration 
the progress and increase  of precision in measuring technology, this can be a 
sign of  possible range variations of $G$ which might be induced 
by some ``new physics"   and/or non-Newtonian 
interactions \cite {Vucetich}, \cite{Melnikov2}. 

In astronomical and geophysical experiments, measurements of possible variations 
of $G$ with cosmic time provide experimental bounds on $\dot{G}/G$, which are 
not very tight. They  span  from $10^{-10}$ to $10^{- 12}$ $yr^{- 1}$depending 
on the experiment and/or observation. Therefore the time variation of $G$ in 
terms of the Hubble parameter $H$ can be written as 
\begin{equation}
\label{scalar-tensor theories of gravity and multidimensional models}
\frac{\dot{G}}{G} = g H,
\end{equation}
where $g$ is a coefficient whose observational bound is $|g| \leq 0.1$.  
A comprehensive and updated discussion  of the various experimental and 
observational constraints on the value of $g$ (as well as on the variation 
of other fundamental ``constants" of nature) has recently been provided 
by Uzan \cite{Uzan}. Therefore, here we will not review the cosmological 
constraints on $g$ imposed by the cosmic microwave background and 
nucleosynthesis. Instead, we will refer the interested 
reader to \cite{Uzan} and references therein. On the other 
hand, there are a number of theories,  such as scalar-tensor theories 
of gravity \cite{Melnikov3} and multidimensional cosmological
 models \cite{Melnikov4}, which lead to  estimates for the variation 
of the gravitational coupling similar to (\ref{scalar-tensor theories
 of gravity and multidimensional models}).  

The introduction of a cosmological term in GR is usually 
considered artificial. Therefore, commonly it is set equal to zero.   
However, the possibility of a  small but non-zero cosmological term is 
suggested  by some recent observational results from Type Ia supernovae 
in distant galaxies and the age problem \cite{Perlmutter}, \cite{Fujii}. 
Present data reveal that the energy density of the field (or quintessence) 
associated with $\Lambda_{(4)}$ exceeds the density of 
ordinary matter \cite{Melnikov2}. There is an extensive literature 
suggesting that the relation 
\begin{equation}
\label{extensive literature}
\Lambda_{(4)} \sim H^2,
\end{equation}
 plays a fundamental role in cosmology \cite{Berman}. This relation 
has been obtained in a number of empirical models \cite{Chakrabarty}, \cite{Wesson}.
 More recently, it was obtained from a model based on the quantum gravitational 
uncertainty principle and the discrete structure of 
spacetime at Plank length \cite{Padmanabhan},  \cite{Vishwakarma}. 
The dependence $\Lambda_{(4)} \sim H^2$ explains the current observations 
successfully and provides a much needed large age of the universe. 

In a recent paper \cite{VarG} we showed that the introduction of a 
time-varying $g_{44}$ in brane-world theory yields a number of cosmological 
models which have good  physical properties but do not admit  constant values 
for $G$ and $\Lambda_{(4)}$. In particular we considered the five-dimensional
 metric  
\begin{equation}
\label{Ponce de Leon solution}
d{\cal S}^2 = 
y^2 dt^2 - t^{2/\alpha}y^{2/(1 - \alpha)}[dr^2 + r^2(d\theta^2 + 
\sin^2\theta d\phi^2)] - \alpha^2(1- \alpha)^{-2} t^2 dy^2,
\end{equation}
where $\alpha$ is a constant. This is a solution of the Einstein 
field equations in five-dimensions \cite{JPdeL 1}. In 
four-dimensions (on the hypersurfaces $y = const.$) it  
corresponds to the $4D$ Friedmann-Robertson-Walker models with 
flat $3D$ sections\footnote{The  energy density $\rho_{eff}$  
and pressure $p_{eff}$ of the effective $4D$ matter satisfy the 
equation of state
$p_{eff} = n \rho_{eff}$,  
where $n = ({2\alpha}/{3} -1)$. Thus  for $\alpha = 2$ we recover 
radiation, for $\alpha = 3/2$  we recover dust, etc.}. 
We used (\ref{Ponce de Leon solution}) as the generating $5D$ space 
for a ${\bf{Z}}_{2}$ symmetric brane-universe filled with perfect fluid, 
with energy density $\rho$ and pressure $p$ satisfying the equation of state 
$p = \gamma \rho$. From Israel's boundary conditions we found that the 
effective $G$ and  $\Lambda_{(4)}$ vary as
\begin{equation}
\label{Var of G from standard cosmology}
\frac{\dot{G}}{G} = - \alpha H, \;\;\;\;\; 8\pi G = 
\frac{k_{(5)}^2[3(\gamma + 1) - \alpha]}{3 (\gamma + 1)}H,
\end{equation}
and  
\begin{equation} 
\Lambda_{(4)} =  \frac{[3(\gamma + 1) - \alpha]^2}{(\gamma + 1 )^2}H^2,
\end{equation}
respectively. Thus, the metric (\ref{Ponce de Leon solution})  
leads an exact  brane-model for a FRW universe where  
where (\ref{scalar-tensor theories of gravity and multidimensional models}) 
and (\ref{extensive literature})  hold.  

A crucial point here is that in brane-world theories the 
quantities $G$ and $\Lambda_{(4)}$ are related to each other through 
the vacuum energy density  of the $3$-brane. They are {\em not} independent, 
 as in Jordan-Brans-Dicke and other multidimensional theories. Thus, 
either both are truly constants or they vary simultaneously, viz.,
\begin{equation}
\label{connection between Lambda dot and G dot}
{\dot{\Lambda}}_{(4)} \sim G \dot{G}.
\end{equation}
This places particular constrains upon the evolution of brane-world models. 
Namely, when we take into account  the experimental
 bounds (\ref{scalar-tensor theories of gravity and multidimensional models}) 
and (\ref{extensive literature}) we obtain an equation for the Hubble parameter which 
has to be solved together with the generalized Friedmann equation on the brane.

Another important feature here is that a cosmological term in  
$5D$ induces   ``natural" constant scales in time, say $\tau_{s}$, 
and length ($l_{s} = c \tau_{s}$) in the physics on the brane.
Therefore, the evolution of the brane universe can be separated 
into three epochs. For $t << \tau_{s}$,  the evolution  of $H$ 
and $\Lambda_{(4)}$   does not differ much from the 
familiar  $H \sim t^{- 1}$ and $\Lambda_{(4)} \sim t^{- 2}$, 
in agreement with the natural dimensions of these quantities 
in $4D$. For $t \sim \tau_{s}$, the overall evolution of the 
expansion scale factor of the brane is dictated by the constant 
scale. For $t >> \tau_{s}$, the evolution depends on whether the 
universe is embedded in a de Sitter or anti de Sitter five-dimensional bulk.

In this work we investigate the consequences  of 
(\ref{scalar-tensor theories of gravity and multidimensional models}) 
and (\ref{extensive literature}) in the framework of brane-world models. 
The natural question to ask is  
whether (\ref{connection between Lambda dot and G dot}) 
could lead to a generic analysis 
of the time variations of $\Lambda_{(4)}$ and $G$. Unfortunately, 
the answer to this question seems to be negative. We will see that 
in order to be able to integrate the equations we need to introduce 
some additional hypothesis  or assumption. 

Regarding $G$, here we consider the simplifying assumption  
that $g $ is a non-vanishing constant (otherwise $G$ and $\Lambda_{(4)}$ are constants). 
Beforehand, we should note that this assumption is {\em not} equivalent to 
requiring  $G \sim H$, in general. The physical meaning of this assumption 
is that the  variation of $g$ is much ``slower" than that of $H$ and $G$, 
namely, $|\dot{g}/g| << |\dot{H}/H|$, $|\dot{g}/g| << |\dot{G}/G|$. Besides, 
the numerical value of $g$ should be small enough as to ensure that one 
obtains the correct abundances, in accordance with the discussion in \cite{Uzan}, 
and not contradict nucleosynthesis.

Regarding $\Lambda_{(4)}$, without any loss of generality we can write
\begin{equation}
\label{assumptions in this work}
\Lambda_{(4)} = \xi(t)H^2,
\end{equation}
where $\xi$ is some function of time. This form of $\Lambda_{(4)}$ is 
convenient for our purposes and is consistent with the fact that 
supernovae data give us  $\Lambda_{(4)} \sim H^2$ at present. In 
general $\xi$ is {\em not} constant, but  a function which can be 
obtained from the equations of the model. We will see that, in  
models with non-vanishing bulk cosmological constant, $\Lambda_{(4)}$ 
is {\em not } strictly proportional 
to $H^2 $, i.e., $\Lambda_{(4)} \neq const \times H^2$. 
These two quantities become  exactly proportional to each other 
only asymptotically in time. This is an interesting feature 
specially in view of latest supernovae results \cite{Tonry} 
suggesting that $\Lambda_{(4)}$ was less important is the 
past and therefore not proportional to $H^2$.

\medskip

The purpose of this work is twofold. Firstly, to  determine the 
general evolution of the expansion scale factor, as well as, the 
variation of $G$, $\Lambda_{(4)}$ and $H$ in a brane-universe which 
is compatible 
with (\ref{scalar-tensor theories of gravity and multidimensional models}) 
and (\ref{extensive literature}).  Secondly, to examine in some detail the 
effects of a  five-dimensional cosmological constant on the evolution of 
the brane universe,  and perhaps on the signature of the extra dimension.

\medskip

Our analysis is rather universal since we make no reference to 
any particular solution of the field equations in $5D$.
For generality, we do not impose the signature of the extra dimension either. 
We consider different possible scenarios where the bulk cosmological constant 
can be zero, positive or negative. For these scenarios, we show that models 
with spacelike extra dimension agree with  the observed accelerating universe. 

The extra dimension can be timelike only if the cosmological constant 
in the bulk is positive. The corresponding cosmological models are well 
behaved and exhibit interesting physical properties, but they do not seem 
to be of much observational significance. 

This paper is organized as follows. In Section $2$ we give a brief summary 
of the theory and of the generalized Friedmann equation on a spatially homogeneous 
and isotropic brane. 
In Section $3$ we show how to incorporate a varying vacuum energy into the scheme. 
We analyze  the compatibility of the generalized Friedmann equation
 with the observational requirements (\ref{scalar-tensor theories of
 gravity and multidimensional models}) and (\ref{extensive literature}). 
We find that the coupled evolution of $G$ and $\Lambda_{(4)}$ completely 
determines the behavior of the brane,  except for some adjustable parameters 
which can be evaluated using some observational data. 
In Sections $4$, $5$ and $6$ we present a detailed study of the behavior 
of the brane-universes under consideration. In Section $7$ we present a 
summary and discussion 

\section{Field equations in $5D$}

In order to facilitate the presentation, and set the notation, we give a 
brief review of the equations in brane-world theory. We consider the metric 
\begin{equation}
\label{metric with g44 not 1}
d{\cal S}^2 = g_{\mu\nu}(x^{\rho}, y)dx^{\mu}dx^{\nu} + \epsilon \Phi^2(x^{\rho}, y) dy^2,
\end{equation} 
where $\epsilon = -1$ or $\epsilon = + 1$ depending on whether the extra dimension 
is spacelike or timelike, respectively. 

 Everywhere we use signature $(+ - - -  \epsilon)$. The Einstein equations in five 
dimensions are
\begin{equation}
\label{field equations in 5D}
{^{(5)}G}_{AB} =
 {^{(5)}R}_{AB} - \frac{1}{2} g_{AB}{^{(5)}R} = {k_{(5)}^2} {^{(5)}T_{AB}}, 
\end{equation}
where $k_{(5)}$ is a constant introduced for dimensional 
considerations and $^{(5)}T_{AB}$ is the five-dimensional 
energy-momentum tensor.  

These equations contain the first and second derivatives 
of the metric with respect to the extra coordinate. These can 
be expressed in terms of geometrical tensors in $4D$.

In absence of off-diagonal terms $(g_{4\mu} = 0)$ the dimensional 
reduction of the five-dimensional equations  is particularly simple \cite{EMT}. 
The usual assumption  is that our spacetime is orthogonal to the extra dimension. 
Thus we introduce the normal unit ($n_{A}n^{A} = \epsilon$) vector, orthogonal to
 hypersurfaces $y = constant$,  
\begin{equation}
n^A = 
\frac{\delta^{A}_{4}}{\Phi}\;  , \;\;\;\;\;  n_{A}= (0, 0, 0, 0, \epsilon \Phi).
\end{equation}
Then, the first partial derivatives can be written 
in terms of the extrinsic curvature 
\begin{equation}
\label{extrinsic curvature}
K_{\alpha\beta} = \frac{1}{2}{\cal{L}}_{n}g_{\alpha\beta} = 
\frac{1}{2\Phi}\frac{\partial{g_{\alpha\beta}}}{\partial y},\;\;\; K_{A4} = 0,
\end{equation}
The second derivatives, $({\partial}^2g_{\mu\nu}/\partial y^2)$, can be 
expressed in terms of the projection $^{(5)}C_{\mu4\nu4}$ of the bulk Weyl 
tensor in five-dimensions, viz.,
\begin{equation}
\label{Weyl tensor}
{^{(5)}C}_{ABCD} = {^{(5)}R}_{ABCD} - \frac{2}{3}({^{(5)}R}_{A[C}g_{D]B} -
 {^{(5)}R}_{B[C}g_{D]A}) + \frac{1}{6}{^{(5)}R}g_{A[C}g_{D]B}.
\end{equation}
The field equations (\ref{field equations in 5D}) can be split up into 
three parts. In terms of the above quantities, the effective field 
equations in $4D$ are,
\begin{eqnarray}
\label{4D Einstein with T and K}
{^{(4)}G}_{\alpha\beta} &=& \frac{2}{3}k_{(5)}^2\left[^{(5)}T_{\alpha\beta} + 
(^{(5)}T^{4}_{4} - \frac{1}{4}{^{(5)}T})g_{\alpha\beta}\right] -\nonumber \\
& &\epsilon\left(K_{\alpha\lambda}K^{\lambda}_{\beta} - 
K_{\lambda}^{\lambda}K_{\alpha\beta}\right) + 
\frac{\epsilon}{2} g_{\alpha\beta}\left(K_{\lambda\rho}K^{\lambda\rho} - 
(K^{\lambda}_{\lambda})^2 \right) - \epsilon E_{\alpha\beta}, 
\end{eqnarray}
where
\begin{eqnarray}
E_{\alpha\beta} &=& {^{(5)}C}_{\alpha A \beta B}n^An^B\nonumber \\
&=& - \frac{1}{\Phi}\frac{\partial K_{\alpha\beta}}{\partial y} +
 K_{\alpha\rho}K^{\rho}_{\beta} - \epsilon \frac{\Phi_{\alpha;\beta}}{\Phi} -
 \epsilon \frac{k^{2}_{(5)}}{3}\left[{^{(5)}T}_{\alpha\beta} + ({^{(5)}T}^{4}_{4} - 
\frac{1}{2}{^{(5)}T})g_{\alpha\beta}\right].
\end{eqnarray}
Since $E_{\mu\nu}$ is traceless, the requirement $E_{\mu}^{\mu} = 0$ gives the 
inhomogeneous wave equation for $\Phi$, viz.,
\begin{equation}
\label{equation for Phi}
{\Phi}^{\mu}_{;\mu} = - \epsilon \frac{\partial K}{\partial y}-
 \Phi (\epsilon K_{\lambda \rho} K^{\lambda \rho} + {^{(5)}R}^{4}_{4}),
\end{equation}
which is equivalent to ${^{(5)}G}_{44} = k^{2}_{(5)}{^{(5)}T_{44}}$ 
from (\ref{field equations in 5D}). The remaining four equations are
\begin{equation}
\label{conservation equation}
D_{\mu}\left(K^{\mu}_{\alpha} - 
\delta^{\mu}_{\alpha}K^{\lambda}_{\lambda}\right) = 
k_{(5)}^2 \frac{{^{(5)}T_{4\alpha}}}{\Phi}.
\end{equation}
In the above expressions, the covariant  derivatives are 
calculated with respect to $g_{\alpha\beta}$, i.e., $Dg_{\alpha\beta} = 0$.

\subsection{The brane-world paradigm}

In the brane-world scenario our space-time is identified with a 
singular hypersurface (or $3$-brane) embedded in an $AdS_{5}$ bulk \cite{RS2}, 
i.e., it is assumed that
the five-dimensional energy-momentum tensor has the form
\begin{equation}
\label{AdS}
{^{(5)}T}_{AB} =  \Lambda_{(5)}g_{AB}, 
\end{equation}
where  $\Lambda_{(5)} < 0$ is the cosmological constant in the bulk. 
For convenience, the coordinate $y$ is chosen such that the 
hypersurface $\Sigma: y = 0$ coincides with the brane. Thus, the 
metric is continuous across $\Sigma$, but the extrinsic curvature $K_{\mu\nu}$ 
is discontinuous. Most brane-world models assume a ${\bf Z}_{2}$ symmetry 
about our brane, namely, 
\begin{eqnarray}
\label{Z2-symmetric metric}
d{\cal S}^2 &=& g_{\mu\nu}(x^{\rho}, + y)dx^{\mu}dx^{\nu} + 
\epsilon \Phi^2(x^{\rho}, + y) dy^2, \;\;\; for\;\; y \geq 0 \nonumber \\
d{\cal S}^2 &=& g_{\mu\nu}(x^{\rho}, - y)dx^{\mu}dx^{\nu} + 
\epsilon \Phi^2(x^{\rho}, - y) dy^2, \;\;\; for\;\;  y \leq 0.
\end{eqnarray}
Thus
\begin{equation}
\label{Z2 symmetry} 
K_{\mu\nu}\mid_{{\Sigma}^{+}} = - K_{\mu\nu}\mid_{{\Sigma}^{-}}.
\end{equation}
Therefore the field equations in the resulting ${\bf Z}_2$-symmetric brane universe 
can be written as 
\begin{equation}
\label{field equations in the Z2 universe}
{^{(5)}\bar{G}}_{AB} =  k^{2}_{(5)}(  \Lambda_{(5)} \bar{g}_{AB}+ 
{^{(5)}\bar{T}}_{AB}^{(brane)}),
\end{equation}
where $\bar{g}_{AB}$ is taking as in (\ref{Z2-symmetric metric}) 
and ${^{(5)}\bar{T}}_{AB}^{(brane)}$, with ${^{(5)}\bar{T}}_{AB}^{(brane)}n^{A} = 0$, 
is the energy-momentum tensor of the matter on the brane
\begin{equation}
\label{energy-momentum tensor in the brane}
{^{(5)}\bar{T}}_{AB}^{(brane)} = 
\delta_{A}^{\mu}\delta_{B}^{\nu}\tau_{\mu\nu}\frac{\delta(y)}{\Phi}.
\end{equation}
The delta function expresses the confinement of matter in the brane, hence
\begin{equation}
\label{brane EMT as result of integration}
\tau_{\mu\nu}(x^{\rho}, 0) = 
\lim_{\xi \rightarrow 0}\int_{- \xi/2}^{\xi/2}{^{(5)}\bar{T}}_{\mu\nu}^{(brane)}\Phi dy.
\end{equation}
From Israel's boundary conditions 
\begin{equation}
\label{boundary conditions}
K_{\mu\nu}\mid_{{\Sigma}^{+}} - K_{\mu\nu}\mid_{{\Sigma}^{-}} = 
- \epsilon k_{(5)}^2 \left({\tau}_{\mu\nu} - \frac{1}{3}g_{\mu\nu}\tau\right),
\end{equation}
 and  the ${\bf Z}_2$ symmetry
\begin{equation}
\label{K in terms of S}
K_{\mu\nu}\mid_{{\Sigma}^{+}} =  - K_{\mu\nu}\mid_{{\Sigma}^{-}} = 
- \frac{\epsilon}{2}k_{(5)}^2 \left({\tau}_{\mu\nu} - \frac{1}{3}g_{\mu\nu}\tau\right),
\end{equation}
we obtain 
\begin{equation}
\label{emt on the brane in terms of K}
\tau_{\mu\nu} = - \frac{2\epsilon}{k_{(5)}^2}\left(K_{\mu\nu} - g_{\mu\nu} K\right).
\end{equation}
Then from (\ref{conservation equation}) and (\ref{AdS}) it follows that
\begin{equation}
\label{conservation of emt on the brane}
\tau^{\mu}_{\nu;\mu} = 0.
\end{equation}
Thus $\tau_{\mu\nu}$ represents the total, vacuum plus matter,  
conserved energy-momentum tensor on the brane. It is 
usually separated in  two parts, 
\begin{equation}
\label{decomposition of tau}
\tau_{\mu\nu} =  \sigma g_{\mu\nu} + T_{\mu\nu},
\end{equation} 
where $\sigma$ is the tension of the brane in  $5D$, which is interpreted 
as the vacuum energy of the brane world, and $T_{\mu\nu}$ represents the 
energy-momentum tensor of ordinary matter in $4D$. 
From (\ref{K in terms of S}), (\ref{emt on the brane in terms of K}) 
and (\ref{decomposition of tau}) we finally get
\begin{equation}
\label{K in terms of matter in the brane}
K_{\mu\nu}\mid_{{\Sigma}^{+}} = -  \frac{\epsilon k_{(5)}^2}{2} \left(T_{\mu\nu} - \frac{1}{3}g_{\mu\nu}(T + \sigma)\right).
\end{equation}
Substituting (\ref{K in terms of matter in the brane}) and (\ref{AdS}) 
into (\ref{4D Einstein with T and K}), we obtain the Einstein equations 
with an {\em effective energy-momentum tensor} in $4D$ 
as \cite{Shiromizu}-\cite{equ. STM-Brane}
\begin{equation}
\label{EMT in brane theory}
^{(4)}G_{\mu\nu} =  {\Lambda}_{(4)}g_{\mu\nu} + 
8\pi G T_{\mu\nu} - \epsilon k_{(5)}^4 \Pi_{\mu\nu} - \epsilon E_{\mu\nu},
\end{equation}
where
\begin{equation}
\label{definition of lambda}
\Lambda_{(4)} = \frac{1}{2}k_{(5)}^2\left(\Lambda_{(5)} - 
\epsilon \frac{ k_{(5)}^2 \sigma^2}{6}\right),
\end{equation}
\begin{equation}
\label{effective gravitational coupling}
8 \pi G =  - \epsilon \frac{k_{(5)}^4 \sigma}{6},
\end{equation}
and\footnote{With this choice of signs, for perfect 
fluid  $\Pi_{\mu \nu} = (1/12)[\rho^2 u_{\mu}u_{\nu}
 + \rho(\rho + 2p) h_{\mu \nu}]$ where $h_{\mu\nu} = u_{\mu}u_{\nu} - g_{\mu\nu}$.}
\begin{equation}
\label{quadratic corrections}
\Pi_{\mu\nu} =  \frac{1}{4} T_{\mu\alpha}T^{\alpha}_{\nu} - 
\frac{1}{12}T T_{\mu\nu} - \frac{1}{8}g_{\mu\nu}T_{\alpha\beta}T^{\alpha\beta} + 
\frac{1}{24}g_{\mu\nu}T^2.
\end{equation}
All these four-dimensional quantities have to be evaluated in 
the limit $y \rightarrow 0^{+}$. The above constitute the 
basic equations in brane-world models. They contain 
higher-dimensional modifications to general relativity. 
Namely, local quadratic energy-momentum corrections via the 
tensor $\Pi_{\mu\nu}$, and the nonlocal effects from the free 
gravitational field in the bulk, transmitted  by $E_{\mu\nu}$.
Another important novel feature is that they provide a working 
definition of the fundamental quantities $\Lambda_{(4)}$ and $G$. 

\subsection{Cosmological settings}

In cosmological applications the five-dimensional metric 
(\ref{metric with g44 not 1}) is commonly taken  in the form
\begin{equation}
\label{cosmological metric}
d{\cal{S}}^2 = n^2(t,y)dt^2 - a^2(t,y)\left[\frac{dr^2}{(1 - kr^2)} +
 r^2(d\theta^2 + \sin^2\theta d\phi^2)\right] + \epsilon \Phi^2(t, y)dy^2,
\end{equation}
where $k = 0, +1, -1$ and $t, r, \theta$ and $\phi$ are the usual 
coordinates for a spacetime with spherically symmetric spatial sections. 

The metric coefficients are subjected to the conditions 
\begin{equation}
\label{conditions for the bulk metric on the brane}
n(t,y)|_{brane} = 1, \;\;\;a(t,y)|_{brane} = a_{0}(t).
\end{equation}
In this way the usual FLRW line element is recovered on the brane 
with $a_{0}$ as scale factor.

The corresponding field equations in the bulk can be written in a very 
compact form in terms of the function $F$, which is a first integral of 
the field equations \cite{Binetruy}, namely, 
\begin{equation}
\label{first integral}
F(t,y) = k a^2 + \frac{(\dot{a}a)^2}{n^2} + \epsilon \frac{(a' a)^2}{\Phi^2},
\end{equation}
where a prime denotes a derivative with respect to $y$. The $(^{0}_{0})$ 
and $(_{4}^{4})$  components  of (\ref{field equations in 5D}) become
\begin{equation}
\label{F prime}
F' = \frac{2a' a^3}{3}k_{(5)}^2 {^{(5)}T}^{0}_{0},
\end{equation}
and
\begin{equation}
\label{F dot}
\dot{F} = \frac{2\dot{a} a^3}{3}k_{(5)}^2 {^{(5)}T}^{4}_{4}.
\end{equation}
Now, the field equations $(^{1}_{1}) = (^{2}_{2}) = (^{3}_{3})$ reduce to
\begin{equation}
\label{first integral in the bulk}
\left(\frac{\dot{a}}{na}\right)^2 =  \frac{k_{(5)}^2 \Lambda_{(5)}}{6} - 
\epsilon \left(\frac{a'}{a \Phi}\right)^2 - \frac{k}{a^2} + \frac{\cal{C}}{a^4},
\end{equation}
where $\cal{C}$ is a constant of integration. 

 The ordinary matter on the brane is usually assumed to be  a perfect fluid 
\begin{equation}
\label{EMT for perfect fluid}
T_{\mu\nu} = (\rho + p)u_{\mu}u_{\nu} - p g_{\mu\nu},
\end{equation}
where the  energy density $\rho$ and pressure $p$ satisfy the isothermal 
equation of state, viz.,
\begin{equation}
\label{equation of state}
p = \gamma \rho, \;\;\;\  0 \leq \gamma \leq 1.
\end{equation}
Thus, the boundary conditions   (\ref{boundary conditions}), 
the ${\bf Z_{2}}$ symmetry (\ref{K in terms of S}), 
and (\ref{K in terms of matter in the brane})  yield \cite{VarG}
\begin{equation}
\label{density}
\rho(t) = \frac{2 \epsilon}{k_{(5)}^2 (\gamma + 1)\Phi|_{brane}}\left[\frac{a'}{a} - 
\frac{n'}{n}\right]_{brane},
\end{equation}
and 
\begin{equation}
\label{tension of the brane}
\sigma =   
\frac{2 \epsilon}{k_{(5)}^2 (\gamma + 1)\Phi|_{brane}}\left[(3\gamma + 2)\frac{a'}{a} + 
\frac{n'}{n} \right]_{brane}.
\end{equation}
These equations show that the tension of the brane and the energy 
density depend on the details of the model. They  enable us to 
investigate the very intriguing  possibility that $\sigma$,  
and consequently $G$ and $\Lambda_{(4)}$, might vary with time. 

Evaluating (\ref{first integral in the bulk}) at the brane we obtain 
the generalized Friedmann equation 
\begin{equation}
\label{generalized FRLW equation}
3\left(\frac{{\dot{a}}_{0}}{a_{0}}\right)^2  =  
\Lambda_{(4)} + 8\pi G \rho - \frac{\epsilon k_{(5)}^4}{12}\rho^2 - \frac{3 k}{a_{0}^2} + 
\frac{3 {\cal{C}}}{a_{0}^{4}}.
\end{equation}
Except for the condition that $n = 1$ at the brane, this 
equation is valid for {\em arbitrary} $\Phi(t,y)$ and $n(t,y)$ in the bulk. 
This equation allows us to examine the evolution of the brane without using 
any particular solution of the five-dimensional field equations. In what 
follows we will omit the subscript $0$.

\section{The equations for varying vacuum energy}

In the above equation $G$ and $\Lambda_{(4)}$ are usually assumed to 
be ``truly" constants. However, as we have already discussed there are 
several models, with reasonable physical properties, for which a variable 
$\Phi$ induces a variation in the vacuum energy $\sigma$ \cite{VarG}. In this 
section we  show how (\ref{generalized FRLW equation}) should be modified as to 
incorporate,  or accommodate, the variation of these fundamental 
physical ``constants", matching observational predictions.  

From (\ref{conservation of emt on the brane}) and (\ref{decomposition of tau}) 
it follows that 
\begin{equation}
\label{Conservation equation}
\sigma_{,\nu} + T^{\mu}_{\nu;\mu} = 0.
\end{equation}
For perfect fluid (\ref{EMT for perfect fluid}) this is equivalent to 
\begin {eqnarray}
\label{conservation equations in explicit form}
\dot{\rho} + (\rho + p)\Theta &=&  - \dot{\sigma}, \nonumber \\
(\rho + p)a_{\nu} + p_{,\lambda}h^{\lambda}_{\nu} &=& \sigma_{,\lambda}h^{\lambda}_{\nu},
\end{eqnarray}
where $\Theta = u^{\mu}_{;\mu}$ is the expansion, 
$a_{\nu} = u_{\nu;\lambda} u^{\lambda}$ is the acceleration 
and $h_{\mu\nu} = u_{\mu}u_{\nu} - g_{\mu\nu}$ is the projector 
onto the spatial surfaces orthogonal to $u_{\mu}$.
In homogeneous cosmological models the second equation above is 
empty; only the first one is relevant. Namely, for the case 
of {\em constant} vacuum energy $\sigma$, and the equation of 
state $p = \gamma \rho$, it  
yields  the familiar relation between the matter energy density 
and the expansion factor $a$, viz., 
\begin{equation}
\label{familiar expansion factor}
\rho \sim \frac{1}{a^{3(\gamma + 1)}}.
\end{equation}
For the case where the vacuum energy is {\em not} constant we need 
some additional assumption. For example, if $\sigma$ is a given 
function $\sigma = \sigma (a)$, then  we integrate the conservation 
equation (\ref{conservation equations in explicit form}) and substitute 
the resulting function $\rho = \rho(a)$ into (\ref{generalized FRLW equation}), 
and thus obtain  the corresponding Friedmann equation. However, we have so far 
no theoretical/observational arguments for the evolution of $\sigma$ in time.

\subsection{$\dot{G}/G \sim H$}

The time variation of $G$ is usually written as $(\dot{G}/G) = g H$, where $g$ is a 
dimensionless parameter. As we have already mentioned in the Introduction, 
nucleosynthesis and the abundance of various elements are used to put 
constraints on $g$. The present observational upper 
bound is $|g| \leq  0.1$ \cite{Uzan}-\cite{Melnikov3}. In what follows we assume 
that $g$ is constant. 

Since $G \sim \sigma$ and $H = \dot{a}/a$ we have $\sigma(a) = f_{0}a^g$, 
where $f_{0}$ is a constant of integration. Thus,   
\begin{equation}
\label{nonhomogeneous equation for the matter density }
\dot{\rho} + 3 \rho(\gamma + 1)\frac{\dot{a}}{a} =  - f_{0}g a^{(g - 1)}\dot{a}.
\end{equation}
First, we consider the case where $\rho$ can be expressed in a way 
similar to (\ref{familiar expansion factor}), i.e., as a power function of $a$. 
Therefore, we find $g = - {3D(\gamma + 1)}/{(f_{0} + D)}$ and 
\begin{equation}
\label{particular but physical solution}
\rho = {D}{a^{ - 3D(\gamma + 1)/(f_{0} + D)}}, 
\end{equation}
where $D$ is a positive constant. In order to simplify the notation we set $f_{0} = F_{0}D$ 
and 
\begin{equation}
\frac{(\gamma + 1)}{(1 + F_{0})} = \beta + 1,
\end{equation}
which is $\beta = (\gamma - F_{0})/(1 + F_{0})$ or $F_{0} = (\gamma - \beta)/(\beta + 1)$.  
With this notation we have
\begin{equation}
\label{matter and vacuum density for a case with vanishing sigma not}
\rho = \frac{D}{a^{3(\beta + 1)}}, \;\;\;  \sigma = 
\frac{D(\gamma - \beta)}{(\beta + 1)a^{3(\beta + 1)}}.
\end{equation}
Notice that $F_{0} \neq 0$ ($\beta \neq \gamma$), otherwise $G = 0$. We also have,
\begin{equation}
\label{ratio G dot over G}
\frac{\dot{G}}{G} = - 3(\beta + 1) H.
\end{equation}
Before going on,  we should become aware of the observational bounds of $\beta$. The lower
 bound comes from the obvious requirement $d\rho/da < 0$, while the upper  one comes from 
the observation that $|g| \leq 0.1$  Thus, 
\begin{equation}
\label{bounds for beta}
-1 < \beta \leq - 0.966, \;\;\;\; \beta = - 0.983 \pm 0.016.
\end{equation}
We will see that $\beta$ is related to $q$ and $\Omega_{\rho}$, the deceleration 
and density parameters, respectively\footnote{Because of the difficulty of reliably 
determining $g$,  the limit set  by (\ref{bounds for beta}) is just exploratory. We use 
this number here, not to make specific observational suggestions, but in order to be able 
to discriminate between what is possible or not.}.

In order to obtain the evolution equation for $a$ we substitute (\ref{matter and vacuum 
density for a case with vanishing sigma not}) into (\ref{generalized FRLW equation}) 
\begin{equation}
\label{Friedmann equation with non-constant sigma}
3\left(\frac{\dot{a}}{a}\right)^2 = \frac{k_{(5)}^2}{2}\Lambda_{(5)}  - 
\frac{\epsilon k_{(5)}^4}{12}\left(\frac{\gamma + 1}
{\beta + 1}\right)^2 \frac{D^2}{a^{6(\beta + 1)}}
-  \frac{3k}{a^2} + \frac{3 {\cal{C}}}{a^{4}}.
 \end{equation}
Let us now introduce the quantities
\begin{eqnarray}
\label{notation}
x &=& a^{3(\beta + 1)}, \nonumber \\
A &=& - \frac{\epsilon k_{(5)}^4}{4}\left(\gamma + 1\right)^2D^2, \nonumber \\
C &=& \frac{3 k_{(5)}^2}{2}\left(\beta + 1\right)^2\Lambda_{(5)},
\end{eqnarray}
in terms of which (\ref{Friedmann equation with non-constant sigma}) becomes
\begin{equation}
\label{equation for x}
\left(\frac{dx}{dt}\right)^2 = 
A + Cx^2 - 9 \left(\beta + 1\right)^2 \left[k x^{(6\beta + 4)/(3\beta + 3)} - 
{\cal{C}}x^{(6\beta + 2)/(3\beta + 3)}\right].
\end{equation}
This equation admits exact integration, in terms of elementary functions, for 
a wide variety  of parameters $A$, $C$, $\beta$ and ${\cal{C}}$. However, we are 
interested in the physical models for which  $\Lambda_{4} \sim H^2$. This introduces 
some additional features which we discuss bellow.

Let us now notice that (\ref{particular but physical solution}) is 
not the general solution of (\ref{nonhomogeneous equation for the matter density }), which
 is
\begin{equation}
\label{general solution for the matter density}
\rho = \frac{{\cal{D}}}{a^{3(\gamma + 1)}} - \frac{f_{0}g}{(3 \gamma + 3 + g)} a^g,
\end{equation}
where ${\cal{D}}$ is the integration constant.  Indeed, setting ${\cal{D}} = 0$ and 
denoting $g = - 3(\beta + 1)$ 
and $D = f_{0}(\beta + 1)/(\gamma - \beta)$, we recover the 
solution (\ref{matter and vacuum density for a case with vanishing sigma not}).
We will see that, for the models under consideration, the observational 
requirement (\ref{bounds for beta}) 
demands ${\cal{D}} = 0$.

\subsection{$\Lambda_{(4)} \sim H^2$}

Without loss of generality we now set  
\begin{equation}
\label{coeff. between lambda and H}
\Lambda_{(4)} = \xi H^2,
\end{equation}
where $\xi$ is an {\em unknown} function of time which we will determine bellow.   Then 
using (\ref{definition of lambda}) and (\ref{matter and vacuum density for a 
case with vanishing sigma not}) we get 
\begin{equation}
\Lambda_{(4)} = \frac{1}{2}k_{(5)}^2\Lambda_{(5)} - 
\epsilon \frac{k_{(5)}^4 D^2 (\gamma - \beta )^2}{12 (\beta + 1)^2 a^{6(\beta + 1)}} =
 \xi \left(\frac{\dot{a}}{a}\right)^2.
\end{equation}
In terms of the notation  (\ref{notation}), this equation becomes
\begin{equation}
\label{the lambda condition}
\left(\frac{dx}{dt}\right)^2 = \frac{3}{\xi}\left[A \frac{(\gamma - \beta)^2}
{(\gamma + 1)^2} + Cx^2\right].
\end{equation}
Thus, equating (\ref{equation for x}) and (\ref{the lambda condition}) we get
\begin{equation}
\label{two important conditions}
A\left[\frac{3(\gamma - \beta)^2}{\xi (\gamma + 1)^2} - 1\right] + 
C\left(\frac{3}{\xi} - 1\right)x^2 + 
9 \left(\beta + 1\right)^2 \left[k x^{(6\beta + 4)/(3\beta + 3)} - 
{\cal{C}}x^{(6\beta + 2)/(3\beta + 3)}\right] = 0.
\end{equation}
This equation should take place for {\em all} $t$ and $x$. Therefore, 
it is important to notice that for the allowed values of $\beta$, given 
by (\ref{bounds for beta}), the powers of $x$ in front of $k$ and ${\cal{C}}$ are 
far from being   zeroth  or  second order. Consequently, 
from (\ref{two important conditions}) we get two compatibility conditions. 
The first one is a condition of physical nature. Namely, 
\begin{equation}
\label{physical conditions}
k = 0, \;\;\; {\cal{C}} = 0.
\end{equation}
Thus, the resulting cosmological models have flat space sections.  
This is compatible with  astrophysical data\footnote{In order to 
avoid misunderstanding, here do not use cosmological parameters
 emerging from  BOOMERANG  or WMAP. The point is that the conditions 
in (\ref{physical conditions}), for a model which does not obey the 
usual Einstein equations, do not contradict well known results obtained 
under the assumption that the (usual) Einstein equations are valid.}  
from BOOMERANG \cite{BOOMERANG} and WMAP \cite{WMAP}. Besides,  the 
constant ${\cal{C}}$, which  is related to the bulk Weyl tensor and corresponds 
to an effective radiation term, is constrained to be small enough at the time of 
nucleosynthesis and it should be negligible today \cite{Langlois}.

The second condition consists of three different options of  
mathematical nature: 
\begin{enumerate}
\item If $A \neq 0$, $C \neq 0$, then (\ref{two important conditions}) 
requires $\xi = 3$ and $(\gamma - \beta)^2 = ( \gamma + 1)^2$;  
\item If $C = 0$, then 
\begin{equation}
\label{second option}
\xi = \frac{3(\gamma - \beta)^2}{(\gamma + 1)^2}. 
\end{equation}
\item $\xi$ is a function of $x$. 
\end{enumerate}
Certainly, the first option  cannot take place by virtue 
of (\ref{bounds for beta});  only the second and the third  
are possible ones. We will see them at work in Sections $4$ and $5$, respectively.

\medskip

If we apply the same arguments 
used above to the general solution (\ref{general solution for the matter density}) 
and  $\sigma = f_{0}a^g$,  instead of (\ref{matter and vacuum density 
for a case with vanishing sigma not}), we get two compatibility conditions. 
The first one for ${\cal{D}} \neq 0$ is very restrictive and  
requires $g = - 3(\gamma + 1)$, which we disregard on observational 
grounds since $|g| \leq 0.1 $. The second condition for ${\cal{D}} = 0$ 
yields the case discussed above. 

\medskip

The conclusion from this section is that the cosmological 
models where $\dot{G}/G \sim H$ and $\Lambda_{(4)} \sim H^2$ have 
flat space sections and zero (or negligible) Weyl radiation from the bulk. 

Their time evolution  follows from 
\begin{equation}
\label{evolution equation}
\int{\frac{dx}{\sqrt{A + Cx^2}}} = (t - t_{0}),
\end{equation}
where $t_{0}$ is a constant of integration.

\medskip

The solution of this integral depends on the constants $A$ and $C$, which 
contain $\epsilon$ and $\Lambda_{(5)}$.  Thus, it allows us to study some 
important questions such as: How does the bulk cosmological constant affect 
the behavior of the brane universe?  Does it influence the choice of the 
signature of the extra dimension?

In the next Sections we analyze the solutions of (\ref{evolution equation}) 
for $\Lambda_{(5)} = 0$, $\Lambda_{(5)} > 0$ and $\Lambda_{(5)} < 0$.

\section{Bulk with $\Lambda_{(5)} = 0$}

We consider first the behavior of the model with vanishing  bulk 
cosmological constant. This case is important because, as we will see later,  
it is the limit  of the  models with $\Lambda_{(5)} \neq 0$ and spacelike signature. 

From (\ref{notation}) it follows that the extra dimension has to be 
spacelike $(\epsilon = -1)$.  The expansion factor is given by
\begin{equation}
\label{a for the simplest model}
a(t) = \left[\frac{k_{(5)}^2}{2}\left(\gamma + 1\right)D t\right]^{1/3(\beta + 1)},
\end{equation}
where we have set $t_{0} = 0$, in such a way that the big bang occurs at $t = 0$. 
Here $\beta$ is related to the deceleration parameter $q = - \ddot{a}a/{\dot{a}}^2$ as
\begin{equation}
\label{q for the simplest model}
q = 2 + 3\beta.
\end{equation}
Therefore, the ratio (\ref{ratio G dot over G}) 
 becomes  $\dot{G}/G = - (q + 1)H$, which is the  same as in 
separable models \cite{VarG}. For the vacuum energy we find
\begin{equation}
\label{vacuum energy in the simplest model}
\sigma = 
\frac{2 \left(\gamma - \beta \right)}{\left(\beta + 1\right)
\left(\gamma + 1\right)k_{(5)}^2 t}.
\end{equation}
Consequently, the $4D$ effective cosmological term $\Lambda_{(4)}$ 
and the gravitational coupling $G$ vary as 
\begin{equation}
\label{condition on gamma minus beta}
\Lambda_{(4)} = \frac{3 (\gamma - \beta)^2}{(\gamma + 1)^2}H^2, 
\;\;\;\;\;\; 8\pi G = \frac{ k_{(5)}^2 (\gamma - \beta)}{\gamma +1} H,
\end{equation}
where $H = \dot{a}/a$ is the Hubble parameter, viz.,
\begin{equation}
\label{Hubble for the simplest model}
H = \frac{1}{3(\beta + 1)} \frac{1}{t}.
\end{equation}
Thus, here the second option (\ref{second option}) is taking place. 
There is a remarkable connection between $\beta$,  $\gamma$ and the 
density parameter $\Omega_{\rho}$. Specifically, 
\begin{equation}
\frac{8 \pi G \rho}{3 H^2} = \Omega_{\rho} = \frac{2(\gamma - \beta)(\beta + 1)}
{(\gamma + 1)^2}.
\end{equation}
Thus, giving $\gamma$ and $\Omega_{\rho}$ we obtain $\beta$, viz., 
\begin{equation}
\label{equation for beta, simplest case}
\left(\beta + 1\right) = 
\frac{1}{2}\left(\gamma + 1\right)\left(1 \pm \sqrt{1 - 2 \Omega_{\rho}}\right),
\end{equation}
which gives an upper bound for $\Omega_{\rho}$. Namely,
\begin{equation}
\label{upper bound for Omega, simplest model}
\Omega_{\rho} \leq \frac{1}{2},
\end{equation}
regardless of the specific value of $\gamma$ and $q$. 

From the above equations, it follows that the density 
parameter $\Omega_{\rho}$ can, in principle, be determined 
by measurements of the deceleration parameter $q$
\begin{equation}
\label{connection between Omega and q}
\Omega_{\rho} = \frac{2(3 \gamma - q + 2)(q + 1)}{9(1 + \gamma)^2}.
\end{equation}
This is a useful feature which results from the assumption 
that $\Lambda_{(5)}$ vanishes. A similar situation occurs in the familiar 
FRW models. We also note the relationship between the density parameters, viz.,
\begin{equation}
\label{relation between density parameters}
\frac{\Lambda_{(4)}}{3 H^2} = \Omega_{\Lambda} = 
\frac{9 \Omega_{\rho}^2(1 + \gamma)^2}{4(1 + q)^2}.
\end{equation} 

\subsection{Behavior of the model for different $\Omega_{\rho}$}

Let us study the behavior of the model for different values of the 
density parameter $\Omega_{\rho}$, within the range of values 
allowed in (\ref{upper bound for Omega, simplest model}). 
From (\ref{a for the simplest model}) and (\ref{equation for beta, simplest case})
 it follows that  
\begin{equation}
\label{FRW1}
a(t) \sim t^{{2}/{3(\gamma + 1)[1 \pm \sqrt{1 - 2 \Omega_{\rho}}]}}.
\end{equation}
Thus, in the upper limit,  for $\Omega_{\rho} \rightarrow 1/2$, 
the evolution of the scale factor of the universe matches the one in  
familiar FRW models, viz., $a(t) \approx t^{2/3(\gamma + 1)}$. Accordingly, 
the deceleration parameter (\ref{q for the simplest model}) 
reduces to $q \approx (1 + 3 \gamma)/2$ which is the usual one 
in FRW cosmologies. An interesting  feature of the model is that, 
in this limit neither the gravitational coupling is constant, nor 
the cosmological term becomes zero. Instead we have
\begin{equation}
\label{FRW2}
8 \pi G = \frac{k_{(5)}^2}{2}H, \;\;\; \Lambda_{(4)} = 
\frac{3}{4}H^2, \;\;\; H = \frac{2}{3(\gamma + 1)}\frac{1}{t},
\end{equation}
where $H$ is the usual Hubble parameter in FRW models, as expected. 

\medskip

Let us now study the behavior of our model  for other values  of $\Omega_{\rho}$. 

\medskip

It is clear that  the deviation from the FRW models increases 
as  $\Omega_{\rho}$ moves away from $1/2$. But  for every value 
of $\Omega_{\rho}$ there are two possible models;  one   for each 
sign in front of the root in (\ref{equation for beta, simplest case}).  They both 
reach the FRW model for $\Omega_{\rho} = 1/2$. The question now is whether they 
satisfy physical conditions.  

Since a reliable and definitive determination of $\Omega_{\rho}$ has 
thus far eluded cosmologists, in our discussion we consider several 
values of $\Omega_{\rho}$, although $\Omega_{\rho} \approx 0.1- 0.3$ 
seem to be the most probably options. 

\subsubsection{Accelerated expansion}

We consider  the behavior of dust-filled universes $(\gamma = 0)$. In 
Table $1$ we illustrate the evolution of such universes corresponding to the 
negative sign in (\ref{equation for beta, simplest case}). It is interesting to 
note that there are no arbitrary parameters or constants in the solution.  
Specifying $\Omega_{\rho}$ we find
 $\beta$, $q$ and $\Omega_{\Lambda} \equiv \Lambda_{(4)}/3H^2$. The 
age of the universe is calculated assuming $H \approx 0.7 \times 10^{- 10} yr^{- 1}$.

\medskip

\begin{center}
\begin{tabular}{|c|c|c|c|c|} \hline
\multicolumn{5}{|c|}{\bf Table 1: $\Lambda_{(5)} = 0$. Accelerated expansion}\\ \hline
 \multicolumn {1}{|c|}{$\Omega_{\rho}$} & 
\multicolumn{1}{|c|}{$\Omega_{\Lambda}$} & $\beta_{(-)}$ & $q_{(-)}$ &
 \multicolumn{1}{|c|}{$({\cal{T}}_{(-)}/10^{10})$ $yr$}\\ \hline\hline
$0.5$ & $0.25$ & $-  0.500$ & $0.500$ & $0.95$\\ \hline
$0.4$ & $0.522$   &  $- 0.723$    & $- 0.170$  & $1.71$ \\ \hline
$0.3$ & $0.664$   &  $- 0.816$    & $- 0.448$   &$2.58$\\ \hline
$0.2$ & $0.783$   &  $- 0.887$    &  $- 0.661$  & $4.21$ \\ \hline
 $0.1$ & $0.889$ & $- 0.947$ & $- 0.841$  & $8.98$ \\ \hline
$0.08$ & $0.907$ & $ - 0.958$  & $- 0.874$ & $11.33$\\ \hline
$0.04$ & $0.936$ & $- 0.979$ & $- 0.938$  & $22.67$\\ \hline
\end{tabular}
\end{center}

This case is interesting because it shows an accelerated expansion 
of the universe, in agreement  with  modern observations. The 
acceleration is driven by the repulsive effect of the ``dark energy" 
associated with  $\Lambda_{(4)}$, which clearly dominates the evolution here. 
We notice that $\Omega_{\rho} + \Omega_{\Lambda} \neq 1$. There is a 
contribution, $\Omega_{\rho^2}$, from the quadratic correction in the 
generalized FRW equation (\ref{generalized FRLW equation}), so 
that $\Omega_{\rho} + \Omega_{\Lambda} + \Omega_{\rho^2} = 1$.  This 
contribution decreases  as the universe gets older. For $\Omega_{\rho} < 0.4$, 
the universe is quite old and  $\Omega_{\rho^2}$ is ``negligible". 

We have already mentioned that for  $\Omega_{\rho} = 0.5$, the expansion scale 
factor is the same as in the de Einstein-de Sitter solution, although the matter 
content is totally different.  We now see that
 $\Omega_{\rho^2} = \Omega_{\Lambda} = 0.25$ for this solution. 
This is another example of the well known fact that in GR the 
same geometry can be attributed to different matter distributions.

Dynamical mass measurements from WMAP Mission reveal that 
the matter content of the universe is about $27 \%$ of the 
critical density ($4 \%$ Atoms, $23\%$ Cold Dark Matter). The 
rest $73 \%$ is usually declared either as ``missing", or Dark 
energy. Also, according to recent measurements\footnote{We would like to 
emphasize that here we are only using the 
measured values of $q$, the rest of the parameters are obtained from the model. }  
the acceleration 
parameter is, roughly, $- 0.5 \pm 0.2$. We note that the entries in 
the third and fourth rows are consistent with this picture of the universe. 
Another remarkable feature here is that the age of the universe is much larger 
than in the usual FRW model (first row).

\subsubsection{Decelerated expansion}

 We now consider the solution with positive sign in front of the 
root in (\ref{equation for beta, simplest case}). For the purpose of 
comparison,  we again consider the evolution of dust universes under  the 
same set of values of $\Omega_{\rho}$ as in the  model with negative sign. 
The relevant parameters are presented in Table $2$.

\medskip

\begin{center} 
\begin{tabular}{|c|c|c|c|c|} \hline
\multicolumn{5}{|c|}{\bf Table 2: $\Lambda_{(5)} = 0$. Decelerated 
expansion}\\ \hline
 \multicolumn {1}{|c|}{$\Omega_{\rho}$} & 
\multicolumn{1}{|c|}{$\Omega_{\Lambda}$} & $\beta_{(+)}$ & $q_{(+)}$ &
 \multicolumn{1}{|c|}{$({\cal{T}}_{(+)}/10^{10})$ $yr$}\\ \hline\hline
$0.5$ & $0.25$ & $-  0.500$ & $0.500$ & $0.95$\\ \hline
$0.4$ & $0.076$   &  $- 0.276$    & $1.170$  & $0.65$ \\ \hline
$0.3$ & $3.37 \times 10^{- 2}$   &  $- 0.184$    & $1.448$   &$0.58$\\ \hline
$0.2$ & $1.27 \times 10^{- 2}$   &  $- 0.113$    &  $1.661$  & $0.53$ \\ \hline
 $0.1$ & $0.37 \times 10^{- 2}$ & $- 0.053$ & $ 1.841$  & $0.50$ \\ \hline
$0.08$ & $0.17 \times 10^{- 2}$ & $ - 0.042$  & $1.874$ & $0.49$\\ \hline
$0.04$ & $0.41 \times 10^{- 3}$ & $ - 0.020$ & $1.938$  & $0.48$\\ \hline
\end{tabular}
\end{center}

\medskip

In the present case the  repulsion associated with the cosmological 
term is negligible. Here the evolution is dominated by the quadratic 
correction term. As an illustration consider the entries  on the third 
row, for which  $\Omega_{\rho^2} \approx 0.7$. The huge gravitational 
attraction produced by this term, for a spacelike extra dimension, explains 
the large deceleration parameter.

This solution presents a number of interesting features. For example, a 
decrease in $\Omega _{\rho}$ entails a decrease in the age of the universe. 
This is the opposite of what we see in Table $1$. Also, ${\cal{T}}_{(+)}$ is 
almost the same for $\Omega_{\rho} \approx 0.1- 0.3$, while in Table $1$ the 
age of the universe changes almost four times in the same range.

\medskip

 However, the  solution with positive sign in (\ref{equation for beta, 
simplest case}) seems to have little in common with present observations. 
Firstly, the universe is quite young.\footnote{This is way the quadratic 
corrections $\Omega_{\rho^2}$ dominate.}  Secondly, it does not fit the 
observational requirements on the ratio $\dot{G}/G$, i.e. on  $\beta$, 
neither on the deceleration parameter.  

\medskip

We would like to finish this Section  by stressing the fact that  
the parameter $\beta$  is related to (i) the ratio $\dot{G}/G$, (ii) 
the deceleration parameter, and (iii) the density parameter $\Omega_{\rho}$ 
through the equations (\ref{bounds for beta}), (\ref{q for the simplest model})
 and (\ref{equation for beta, simplest case}), respectively.

\section{de Sitter bulk, $\Lambda_{(5)} > 0$} 

Although in brane-world theory our universe is embedded in a 
higher-dimensional space  with negative cosmological constant, 
the solutions to  the evolution equation (\ref{evolution equation}) 
depend analytically on $\Lambda_{(5)}$, allowing us to continue to
 the range of positive bulk cosmological constant. In this Section
 we explore the case with $\Lambda_{(5)} > 0$. An attractive feature 
of this case is that the extra dimension can be either spacelike or 
timelike. We will discuss these two cases separately. 

\subsection{Spacelike extra dimension $\epsilon = -1$}

The integration of (\ref{evolution equation}) yields
\begin{equation}
a^{3(\beta + 1)} = \frac{k_{(5)}(\gamma + 1) D}
{(\beta + 1)\sqrt{6 \Lambda_{(5)}}}\sinh{\sqrt{C}t},
\end{equation}
where once more we have set $t_{0} = 0$ and 
\begin{equation}
\label{C for the positive Lambda solution}
\sqrt{C} = k_{(5)}\left(\beta + 1\right)\sqrt{\frac{3 \Lambda_{(5)}}{2}}.
\end{equation}
The deceleration parameter here is 
\begin{equation}
\label{q for 3.1.2}
q = 2 + 3\beta - 3\left(\beta + 1\right)\tanh^{2}{\sqrt{C}t}.
\end{equation}
The matter and vacuum density vary as 
\begin{equation}
\label{matter and vacuum density for 3.1.2}
\rho = \frac{2 \sqrt{C}}{k_{(5)}^2\left(\gamma + 
1\right)\sinh{\sqrt{C}t}}, \;\;\;\;\;\sigma = \frac{2 \left(\gamma - 
\beta\right)\sqrt{C}}{k_{(5)}^2\left(\gamma + 
1\right)\left(\beta + 1\right){\sinh{\sqrt{C}t}}}.
\end{equation}
As in the previous  solution we have $\dot{G}/G = - 3(\beta + 1) H$ and 
\begin{equation}
\Lambda_{(4)} = 
\frac{3 \left(\gamma - \beta\right)^2}{\left(\gamma + 
1\right)^2}H^2 + \frac{k_{(5)}^2 \Lambda_{(5)}}{2}\frac{\left[\left(\gamma + 1\right)^2 - 
\left(\gamma - \beta\right)^2\right]}{\left(\gamma + 1\right)^2},
\end{equation}
\begin{equation}
\label{G 3.1.2}
8 \pi G = 
\frac{ k_{(5)}^2\left(\gamma - \beta\right)}{\left(\gamma + 1\right)}H \sqrt{1 - 
\frac{k_{(5)}^2 \Lambda_{(5)}}{6 H^2}},
\end{equation} 
where 
\begin{equation}
\label{H for 3.1.2}
H = \frac{\sqrt{C}}{3\left(\beta + 1\right)\tanh{\sqrt{C}t}}.
\end{equation}
We note that $\Lambda_{(4)}$ is always positive and $H$ is a 
monotonically decreasing function of time bounded bellow by 
$H =  k_{(5)}\sqrt{\Lambda_{(5)}/6}$. This assures the positivity of $G$, 
as long as $\gamma > \beta$.  

The general behavior of the solution is as follows
\begin{enumerate}
\item For $\Lambda_{(5)} =  0$ we recover the previous solution. 
\item For small values of  $t$, near the big bang, this model behaves 
exactly as the previous one 
(\ref{a for the simplest model})-(\ref{Hubble for the simplest model}). 
\item At large times the universe is expanding with positive acceleration, 
$q \approx - 1$. The expansion is exponential and $\Lambda_{(4)}$ 
tends to  $\tilde{\Lambda} \equiv k_{(5)}^2 \Lambda_{(5)}/2$.
\end{enumerate}

Now, if we combine (\ref{q for 3.1.2}) and (\ref{H for 3.1.2}) we obtain 
\begin{equation}
\label{C for 3.1.2}
C = 3H^2 \left(2 + 3 \beta - q\right)\left(1 + \beta\right).
\end{equation}
Using this expression into $8 \pi G \rho$ 
from (\ref{matter and vacuum density for 3.1.2}) we get
\begin{equation}
\label{beta for solution 4.1}
\beta = \gamma -\frac{3 \Omega_{\rho} (\gamma + 1)^2}{2(1 + q)}.
\end{equation}
The condition $(\beta + 1) > 0$ sets an upper limit on the density parameter, viz., 
\begin{equation}
\label{condition for Omega, exponentially expanding sol.}
\Omega_{\rho} < \frac{2(1 + q)}{3(1 + \gamma)}.
\end{equation}
We also obtain an interesting expressions for $\Lambda_{(4)}$. Namely,
\begin{equation}
\label{varying Lambda 4, for dS bulk}
\Lambda_{(4)} = \left[\frac{(\gamma - \beta)^2(1 + q) + 
(1 + \gamma)^2(2 + 3 \beta - q)}{(1 + \beta)(1 + \gamma)^2}\right] H^2.
\end{equation}
Coming back to (\ref{coeff. between lambda and H}), here the factor ($\xi$) 
in front of $H^2$ is {\em not} constant, because $q$ is a function of time. 
Thus, this solution corresponds to the third option mentioned 
after (\ref{second option}).

For the gravitational ``constant" we get
\begin{equation}
\label{gravitational constant}
8 \pi G = 
k_{(5)}^2 \frac{(\gamma -\beta)\sqrt{1 + q}}{\sqrt{3}(\gamma + 1)\sqrt{1 + 
\beta}}H
\end{equation}
The above equations (\ref{beta for solution 4.1}), (\ref{varying Lambda 4, for dS bulk})
 relate the observational quantities $\gamma$, $\Omega$, $q$ and  $\beta $.  The age 
of the universe is 
\begin{equation}
{\cal{T}} = 
\frac{1}{\sqrt{C}}\tanh^{- 1}\left(\frac{2 + 3\beta - q}{3 (\beta + 1)}\right)^{1/2},
\end{equation}
where $C$ is obtained from (\ref{C for 3.1.2}). Finally, the five-dimensional 
quantities $k_{(5)}$ and $\Lambda_{(5)}$ can be evaluated 
from (\ref{gravitational constant}) and (\ref{C for the positive Lambda solution}). 
We note that there are no free parameters left in the solution.

\subsubsection{Characteristic time}

It is important to notice that a non-vanishing cosmological constant 
in the bulk induces a natural time scale, in $4D$. We define it 
as $\tau_{s} = \sqrt{3/\tilde{\Lambda}}$, 
where $\tilde{\Lambda} \equiv k_{(5)}^2 \Lambda_{(5)}/2$. 
Thus, $\tau_{s} = 3(\beta + 1)/\sqrt{C}$ or, in terms of observational quantities 
\begin{equation}
\label{characteristic time}
\tau_{s} = \frac{1}{H}\left[\frac{4(1 + q)^2 - 
9 \Omega_{\rho}^2(1 + \gamma)^2}{4(1 + q)^2 \Omega_{\Lambda} - 
9 \Omega_{\rho}^2 (1 + \gamma)^2}\right]^{1/2}.
\end{equation}
We remark that although the quantities on the r.h.s. are 
functions of time, $\tau_{s}$ is a ``universal" constant fixed 
by the five-dimensional embedding bulk. This equation shows how we 
can evaluate this constant from measurements performed in $4D$.
The denominator vanishes and $\tau_{s} \rightarrow \infty$ when the 
density parameters are related as in (\ref{relation between density parameters}). 
Thus, the ``upper" bound for $\tau_{s}$ takes place in the models discussed in the 
previous Section. 

In order  to get an expression for the lower bound, we 
rewrite (\ref{characteristic time}) as 
\begin{equation}
\label{evaluation of characteristic time}
\frac{4(1 + q)^2}{9 \Omega_{\rho}^2 (1 + \gamma)^2} = 
\frac{H^2 \tau_{s}^2 - 1}{\Omega_{\Lambda} H^2 \tau_{s}^2 - 1}.
\end{equation}
From (\ref{beta for solution 4.1}) it follows that the numerator 
in (\ref{characteristic time}) is always positive. Then, from $\Omega_{\Lambda} < 1$ 
we obtain $(H^2 \tau_{s}^2 - 1) > 0$. Consequently from (\ref{evaluation of 
characteristic time}) we get $(\Omega_{\Lambda}H^2 \tau_{s}^2 - 1) > 0$. Thus 
\begin{equation}
\label{lower bound for characteristic time}
\frac{1}{H \sqrt{\Omega_{\Lambda}}} < \tau_{s} < \infty
\end{equation}
For a universe with  $\Omega_{\Lambda} \approx  0.7$ 
and $H \approx 0.7 \times 10^{- 10}$ $yr^{- 1}$, the lower bound for the 
characteristic scale is $\approx 1.7 \times 10^{10}$ years or $17$ billion 
years, which is more that the age of the universe according to the data from WMAP Mission.

\subsubsection{Observational constrains}

Let us now use the observational bound (\ref{bounds for beta}).  
From (\ref{beta for solution 4.1})  we obtain
\begin{equation}
\label{general restrictions}
\frac{3 \Omega_{\rho}(1 + \gamma)}{2} - 1 < q < \frac{3\Omega_{\rho}(1 + \gamma)^2}{2 (\gamma + 0.966)} - 1,
\end{equation}
where we have assumed $(\gamma + 0.966) > 0$. This equation 
implies, $q < 0$, an  accelerated expansion of the universe  in the 
range of values allowed for $\Omega_{\rho}$. However, not all $q$ and 
$\Omega_{\rho}$ generate adequate physical models. We have to take into 
account that here $\Lambda_{(5)} > 0$. Therefore, from (\ref{notation}) it 
follows that $C > 0$. Consequently,  (\ref{C for 3.1.2}) 
requires  $(2 + 3 \beta - q) > 0$. If we apply the observational 
bounds (\ref{bounds for beta}),  as well as (\ref{condition for Omega, 
exponentially expanding sol.}), we get
\begin{equation}
\label{strong requirement}
- 1 < q < - 0.88, \;\;\  \Omega_{\rho} \leq \frac{0.08}{(1 + \gamma)}, 
\end{equation} 
which are even more restrictive than (\ref{general restrictions}).  
According to the present estimates for $q$ and $\Omega_{\rho}$ we are 
nowhere near these values. Thus,  although the present model is 
attractive from a theoretical point of view, it seems to have limited 
practical application.

\subsection{Timelike extra dimension $\epsilon = + 1$}

  For completeness we now consider the case where the fifth 
dimension is timelike, although these models are controversial in 
many respects, including the problems of causality and quantization. We 
are not going to discuss these problems here, instead we refer the interested 
reader to \cite{BMC} and references therein. 

In this case the scale factor is given by  
\begin{equation}
a^{3(\beta + 1)} = 
\frac{k_{(5)}(\gamma + 1) D}{(\beta + 1)\sqrt{6 \Lambda_{(5)}}}\cosh{\sqrt{C}t},
\end{equation}
where once more we have set $t_{0} = 0$.  The expressions for the 
physical quantities $\rho$, $\sigma$, $G$, $\Lambda_{(4)}$ and $H$ are 
formally obtained from the solution discussed in Section  $5.1$  by 
changing $\sinh\sqrt{C}t \rightarrow \cosh\sqrt{C}t$ and 
$\tanh \sqrt{C}t \rightarrow 1/\tanh\sqrt{C}t$.  However, there 
is a qualitative difference between the solutions.
\begin{enumerate} 
\item This is a ``bouncing" solution where the universe never collapses 
to a singularity. The universe is always expanding with positive acceleration. 
\item The Hubble parameter is an {\em increasing} function of time and 
is bounded above by $k_{(5)}\sqrt{\Lambda_{(5)}/6}$. 
\item The vacuum energy must be negative in order to ensure the positiveness 
of $G$. Namely, 
\begin{equation}
G = 
\frac{k_{(5)}^2 (\beta - \gamma)\sqrt{C}}{3(\gamma + 1)(\beta + 1)\cosh{\sqrt{C}t}}, 
\;\;\; \beta > \gamma
\end{equation}
\end{enumerate}

\medskip

We conclude that the model with a timelike extra dimension seems to be 
of no observational significance because for all $t$ we have $q < -1$, 
which is the opposite to what we expect $(q > -1)$,  based on  recent 
measurements.

\subsection{Asymptotic behavior and classical inflation}

The late time behavior of solutions with $\Lambda_{(5)} > 0$ is similar 
for both signatures, $\epsilon =  - 1$ or $\epsilon = + 1$, although they 
drastically differ at the origin of the universe. In both cases the expansion 
factor becomes identical to the one in the de Sitter solution, 
\begin{equation}
\label{asymptotic de Sitter}
a(t) \sim \exp{\sqrt{(\tilde{\Lambda}/3)}t} = \exp({t/\tau_{s}}),
\end{equation}
 as $t >> \tau_{s}$. It is clear that we are far from this asymptotic regime. 
It should be emphasized that this exponential behavior {\em does not} arise 
from a ``false-vacuum" equation of state $p = -  \rho$ as in inflation, 
because $\gamma \neq -1$ in these solutions. The reason for this is that 
the assumption $\dot{G}/G = gH$ is equivalent to the requirement that $\rho$ 
and $\sigma$ form a combined fluid with energy density $\bar{\rho} = \rho + \sigma$ 
and pressure $\bar{p} = \beta \bar{\rho}$. Then, the observational constraint $-1 < \beta < - 0.966$ 
implies that the combined fluid behaves nearly like a cosmological constant, which 
dominates at late times and thus producing inflation. 

\subsubsection{Vacuum equation of state}

For consistency, we should now show that this model is compatible with 
classical inflation for the equation of state of false-vacuum,   $\gamma = -1$.  
Indeed, if $\gamma = -1$, then from (\ref{notation}) it follows that $A = 0$. 
Thus, the evolution equation (\ref{evolution equation}) requires $C > 0$, 
which entails   $\Lambda_{(5)} > 0$.  Integrating (\ref{evolution equation})
 we get $x = \exp{\sqrt{C}(t - t_{0})}$, which in terms  of the  original 
notation is identical to the de Sitter expression (\ref{asymptotic de Sitter}).   

 \medskip

We conclude this section with the following comments.

\medskip

$(i)$  From (\ref{condition for Omega, exponentially expanding sol.}) it 
follows that for a every given value of $q$, there is a  {\em range} of possible 
values for $\Omega_{\rho}$. Similarly, for a given  $\Omega_{\rho}$, there is a  
range of allowed values for $q$, which is given by (\ref{general restrictions}). 
This is different from the case with $\Lambda_{(5)} = 0$ where they are related
 by (\ref{connection between Omega and q}). Therefore,  
for $\Lambda_{(5)} \neq 0$ the sole specification of one 
of these parameters is not enough to determine the characteristics 
of the model. Here we have to specify both, independently. This 
additional degree of freedom is a consequence of the introduction of a 
non-vanishing cosmological constant in the bulk. 

$(ii)$ Measurements in $4D$ allow to ``predict" the value of the five-dimensional 
constants $k_{(5)}$ and $\Lambda_{(5)}$.  Namely, if we measure $(\Omega_{\rho}, q)$, 
then $\beta$ follows from  (\ref{beta for solution 4.1}). 
The value of $\Omega_{\Lambda}$ is then  obtained 
from (\ref{varying Lambda 4, for dS bulk}). Finally, 
from (\ref{gravitational constant}) and (\ref{characteristic time}) 
we evaluate  the $5D$ constants $k_{(5)}^2$ and $\tau_{s}$ (and/or $\Lambda_{(5)}$)
 in terms of $H^{-1 }$. If we have $(\Omega_{\rho}, \Omega_{\Lambda})$, 
then $\beta$ and $q$ are given by the solutions of the system of 
equations (\ref{beta for solution 4.1}) and (\ref{varying Lambda 4, for dS bulk}), 
then following the same steps  as above we get the rest of the parameters. 

\section{Anti-de Sitter bulk, $\Lambda_{(5)} < 0$}

This case is important, it corresponds to the brane world scenario where our 
universe is identified with a singular hypersurface (or a three-brane) embedded 
in an $AdS_{5}$ bulk.

 For $\Lambda_{(5)} < 0$ the extra dimension has to be spacelike and the evolution 
of the scale factor is given by
\begin{equation}
\label{recollapsing model}
a^{3(\beta + 1)} = 
\frac{k_{(5)}(\gamma + 1) D}{(\beta + 1)\sqrt{6 |\Lambda_{(5)}|}}\sin{\sqrt{|C|}t},
\end{equation}
where once more we have set $t_{0} = 0$ and 
\begin{equation}
\sqrt{|C|} = k_{(5)}\left(\beta + 1\right)\sqrt{\frac{3 |\Lambda_{(5)}|}{2}}.
\end{equation}
In the present case the  deceleration parameter is 
\begin{equation}
q = 2 + 3\beta + 3\left(\beta + 1\right)\tan^{2}{\sqrt{|C|}t},
\end{equation}
where 
\begin{equation}
|C| = 3 H^2 (q -2 - 3 \beta)(1 + \beta).
\end{equation}

This model is formally obtained from the solution in Section $5.1$ by making 
the change $\sqrt{\Lambda_{(5)}} \rightarrow i\sqrt{|\Lambda_{(5)}|} $. 
Thus $\sqrt{C} \rightarrow i\sqrt{|C|}$. 
Consequently,  $\left(\sinh{\sqrt{C}t}\right)/i \rightarrow \sin{\sqrt{|C|}t}$. 
However, they are drastically different from each other. 

The  present solution $(\Lambda_{(5)} < 0)$ represents a spatially flat
 but {\em recollapsing} universe. The recollapse  time ${\cal{T}}_{rec}$ is 
given by $\sin\sqrt{|C|}{\cal{T}}_{rec} = 0$, which  in terms of the 
characteristic time $\tau_{s}$ defined in Section $5.1.1$ 
becomes\footnote{In this case we use the magnitude of $\Lambda_{(5)}$, 
namely, $\tilde{\Lambda} = k_{(5)}^2 |\Lambda_{(5)}|/2$. Therefore, for 
the evaluation of $\tau_{s}$ in (\ref{characteristic time}) we have to  
take the magnitude of the quantity in parenthesis, otherwise $\tau_{s}$ 
would be a complex number.}
\begin{equation}
{\cal{T}}_{rec} = \frac{\pi \tau_{s}}{6 (\beta + 1)}.
\end{equation} 
We note that for the physical values (\ref{bounds for beta}), 
${\cal{T}}_{rec} > 15 \tau_{s}$. The age of the universe can be 
written as 
\begin{equation}
{\cal{T}} = 
\frac{2{\cal{T}}_{rec}}{\pi}\tan^{- 1}\left(\frac{q - 2 - 3 \beta}{3 (\beta + 1)}\right)^{1/2}.
\end{equation}
In Table $3$ we use this solution to estimate the 
parameters $\Omega_{\Lambda}$, $q$, in dust-filled universes, 
with various values of $\Omega_{\rho}$. We also evaluate, 
in units of $H^{-1}$, the characteristic and recollapse time, 
$\tau_{s}$ and ${\cal{T}}_{rec}$, as well as the age of the universe  ${\cal{T}}$. 
\begin{center}
\begin{tabular}{|c|c|c|c|c|c|c|} \hline
\multicolumn{6}{|c|}{\bf Table 3: $AdS$ bulk, $\Lambda_{(5)} < 0$}\\ \hline
 \multicolumn {1}{|c|}{$\Omega_{\rho}$} & $\Omega_{(\Lambda)}$& $q$ & 
$H \tau_{s}$ & $H {\cal{T}}_{rec}$ &  
 \multicolumn{1}{|c|}{$H {\cal{T}}$} \\ \hline\hline
$0.5$ & $0.495$& $- 0.236 \pm 0.013$ & $0.278$& $7.943$&$6.569$    \\ \hline
$0.4$ & $0.596$& $- 0.389 \pm 0.011$   &$0.311$& $9.044$&$7.650$   \\ \hline
$0.3$ &$0.697$ & $- 0.542 \pm 0.008$   & $0.359$& $10.760$&$8.396$   \\ \hline
$0.2$ &$0.798$ & $- 0.694 \pm 0.006$  & $0.487$ &$13.004$&$9.246$   \\ \hline
 $0.1$ &$0.898$ & $- 0.846 \pm 0.003$ & $1.005$ &$20.254$&$10.048$   \\ \hline
$0.08$ &$0.919$ & $- 0.877 \pm 0.002$ & $1.211$ &$25.996$&$11.417$  \\ \hline
\end{tabular}
\end{center}
For a given $\Omega_{\rho}$ the values of  $q$,  as obtained 
from (\ref{general restrictions}), are  spread over a ``small" range. 
In other words, if we know $\Omega_{\rho}$, we get $q$ with a great  accuracy, 
which is  up to  $1 \% - 5\%$. We should emphasize that this precision comes from 
the experimental bounds on $\dot{G}/G$. Then, we use the  
mean value of $q$ to obtain $\beta$ from (\ref{beta for solution 4.1}), 
which we substitute into (\ref{varying Lambda 4, for dS bulk}) 
to get $\Omega_{\Lambda}$.  

We note that the entries on rows two, three and four are similar 
to those in Table $1$. 
Here the deceleration parameter $q$ is a monotonically increasing 
function of time and changes its sign for $\beta$ less than 
$ \approx - 0.66$. Therefore, for the values allowed by  
(\ref{bounds for beta}), the universe initially expands 
with acceleration $(q < 0)$, then changes sign becoming 
positive and $q \rightarrow + \infty$ for $t \rightarrow {\cal{T}}_{rec}$. 

This behavior is a consequence of the effective cosmological term $\Lambda_{(4)}$ 
which is positive near the big bang and, as $t \rightarrow {\cal{T}}_{rec}$, becomes 
\begin{equation}
\Lambda_{(4)} \rightarrow 
\frac{k_{(5)}^2 \Lambda_{(5)}}{2}\left[1 - \frac{(\gamma - \beta)^2}{(\gamma + 1)^2}\right],
\end{equation}
which is negative for $\Lambda_{(5)} < 0$.

\medskip

Finally, we would like to note some similarities between solutions with a 
spacelike extra dimension $\epsilon = - 1$. Firstly, they have an analogous 
behavior near the big bang, regardless of the value of the bulk cosmological 
constant. Secondly, the gravitational coupling of geometry to matter, $G$,  is 
divergent for $t \rightarrow 0$.
This  last feature is of particular interest,  it suggests that the 
gravitational interaction was much stronger in the past than at the present time.

\section{Summary and conclusions}

In summary, we have studied the consequences of  the conditions 
$\dot{G}/G = g H$, with $|g| \leq 0.1$, and $ \Lambda_{(4)} \sim H^2$ on 
cosmological models based on the brane-world scenario. These $two$ conditions 
lead to the requirement $k = 0$, ${\cal{C}} = 0$, in (\ref{physical conditions}). 
The same physics is obtained if we  assume $\dot{G}/G = g H$  and $k = 0$, ${\cal{C}} = 0$. 
In this case $ \Lambda_{(4)} \sim H^2$ is {\em not} an extra condition   
but a consequence of these assumptions. 
However, if we choose to postulate 
$ \Lambda_{(4)} \sim H^2$, {\em  and} $k = 0$, ${\cal{C}} = 0$, 
then we do not obtain $\dot{G}/G = g H$, except for the case 
where $\Lambda_{(5)} = 0$. In other words, we recover only the 
models discussed in Section $4$.

The study of these cosmological models is attractive for  several reasons.
 Firstly, conditions (\ref{scalar-tensor theories of gravity and multidimensional models}) 
and (\ref{extensive literature}) are motivated by a number of theoretical and experimental 
observations. Secondly, brane theory provides a framework within which to consider 
the {\em simultaneous} variation of $G$ and $\Lambda_{(4)}$, because they are related 
to  the vacuum energy, which in principle could vary with time. Thirdly, the theory 
has no arbitrary parameters. 

Although we introduce two ``new" parameters, besides the observed 
$g$ (or $\beta$),  viz.,  $\Lambda_{(5)}$ and $k_{(5)}$, they can be 
evaluated by means of quantities measured in $4D$ only. Specifically, 
for $\Lambda_{(5)}$ we have  
\begin{equation}
\Lambda_{(5)} = 
\frac{(2 + 3\beta - q)(\gamma - \beta)\sqrt{1 + q}}{\sqrt{3}(\gamma + 1)
(1 + \beta)^{3/2}}\left(\frac{H^3}{4 \pi G}\right).
\end{equation}
As an illustration we consider the entries on the third row in Table $3$, 
for which we get $\Lambda_{(5)} \approx - 8.8 \times 10^{- 46}$ $gr$ $cm^{- 3}$ $s^{- 1}$. 
The value of $k_{(5)}^2$ can be obtained from (\ref{gravitational constant}) 
as $k_{(5)}^2 \approx 26.23 \times 10^{10}$ $gr^{- 1}$$cm^3$ $s^{- 1}$. Thus, 
the r.h.s. of the field equations in (\ref{field equations in 5D}) is 
a very ``small" 
number, viz., $k_{(5)}^2\Lambda_{(5)} \approx 2.3 \times 10^{- 34}$ $s^{- 2}$. 
In terms of the Plank units of time $t_{Plank} = 10^{- 43}$ $s$, 
we get $k_{(5)}^2\Lambda_{(5)}/2 \sim |\Lambda_{(4)}| \approx 10^{- 120}$ $t_{Plank}^{- 2}$, 
as one expected. Although this number is small, it has a crucial influence on the 
global behavior of the universe.

The model requires the universe to have flat $(k = 0)$ space 
sections, which is consistent with observations. Also, it is 
interesting that all possible scenarios, except those in Table $2$, 
agree with the observed  accelerated expansion of the universe and are 
dominated by the Dark energy associated with $\Lambda_{(4)}$. 

In addition, the observed value of $q = - 0.5 \pm 0.2$ fits the ones 
predicted by the models with $\Lambda_{(5)} \leq 0$, as shown  in 
Tables $1$ and $3$ for  $\Omega_{\rho} \approx 0.1 - 0.3$. This is 
different from the models with positive bulk cosmological constant. 
These present good physical properties but do not appear to be compatible 
with current observations, regardless of the signature of the extra dimension.   

In all cases the cosmological models with spacelike extra dimension 
have similar behavior near
the big-bang. If $\Lambda_{(5)} = 0$, then the induced cosmological term 
in $4D$ is positive and the brane universe expands continuously in a power-law 
time-dependence. If $\Lambda_{(5)} > 0$, then the universe becomes 
dominated by a positive cosmological term $\Lambda_{(4)}$,  which tends 
to a constant value. The effect is an asymptotic de Sitter expansion, 
which occurs without a false-vacuum equation of state and regardless of 
the signature of the extra dimension. If $\Lambda_{(5)} < 0$, 
then $\Lambda_{(4)}$ is initially positive, but later changes its sign. 
The universe becomes dominated by a negative cosmological term that causes 
the universe to recollapse.

\medskip

 Consequently, an inevitable conclusion of our work is that the 
universe must recollapse at some time in the future if it is  
embedded in an Anti-de Sitter five-dimensional bulk, which is the 
usual case in brane models. Fortunately, according to Table $3$, we 
are nowhere near the time of recollapse. We emphasize that this behavior 
is independent on the (small) size of the (negative) bulk cosmological constant. 

\medskip

However, we note that the entries in Tables $1$ and $3$ are not very 
different from each other for $\Omega_{\rho} \approx 0.1 - 0.3$. This is 
explained by the ``negligible" difference  between the numbers 
in the r.h.s. of the  field equations (\ref{field equations in 5D}) 
in both cases.  The entries in Table $1$ are calculated 
for $k_{(5)}^2\Lambda_{(5)} = 0$, while the ones in 
Table $3$ for $k_{(5)}^2\Lambda_{(5)} \approx 2.3 \times 10^{- 120}$ $t_{Plank}^{-2}$. 
Such a small difference does not significantly alter the evolution of the universe.

Therefore, the solution with accelerated expansion  of Section $4.1.1$ 
gives  a good description, to a first order approximation,  of the 
evolution of a $4D$ universe embedded in a five-dimensional spacetime 
with non-vanishing $\Lambda_{(5)}$ and spacelike extra dimension.
We would like to emphasize that the whole analysis  in this paper is 
independent of any particular solution used in the five-dimensional bulk. 
This is a great virtue of brane-world models as noted at the end of Section $2$.

However, one could still ask whether the brane metrics described here 
can be embedded in a five-dimensional bulk. The answer to this question 
is positive. Clearly the model with $\Lambda_{(5)} = 0$, which generalizes 
the familiar power-law solution characteristic of flat FRW universes, is 
embedded in the cosmological metric  with separation of 
variables (\ref{Ponce de Leon solution}) which is 
discussed in \cite{VarG}. Regarding the models 
with $\Lambda_{(5)} \neq 0$, they can be embedded in 
five-dimensional ``wave-like" cosmologies of the type 
discussed in \cite{BMC}. If in equation $(38)$ of \cite{BMC} 
we take variable $\sigma$ as here in (\ref{matter and vacuum 
density for a case with vanishing sigma not}), then the scale 
factor $a$ for such wave-like models is governed by an equation 
which is identical to (\ref{Friedmann equation with non-constant sigma}) 
in this paper.

We would like to finish with the following comments. The issue 
of constraining brane universes with observational data
like supernova, CMB and quasars has been done in detail for brane
cosmology in \cite{Pam1} and \cite{Pam2}.  Our aim in this work has 
been to study the simultaneous variation of  $G$
and $\Lambda_{(4)}$ and constraining the model with observations. Such study
 has not been
done before.  Besides we have very little indication of the

 evolution  of $G$ and $\Lambda_{(4)}$ in time. Our working hypothesis, 
that  $\dot{G}/G = g H$, is an extrapolation  of the present limit of $|\dot{G}/G |$. 
It is not clear whether it remains valid as early as at nucleosynthesis. Many other 
important questions remain open. Among them the behavior of perturbations and  
structure formation. Also the issues of obtaining the $4$-dimensional Newton law as 
well as the quantization in models with positive bulk cosmological constant and 
models with timelike extra dimension, respectively.  These questions are beyond the 
scope of the present work.

\end{document}